\documentclass[aps,reprint,prl,superscriptaddress,showpacs,floats,floatfix]{revtex4-2}
\usepackage{graphicx}
\usepackage{amsmath,graphicx,dcolumn}
\usepackage[usenames]{color}
\usepackage{datetime}
\usepackage{textcomp}
\usepackage{ulem}
\usepackage{morefloats}
\usepackage{mathptmx}
\usepackage{indentfirst}
\usepackage{mathrsfs}
\usepackage{bm}
\usepackage{siunitx}
\usepackage{ulem}
\usepackage[colorlinks=true, linkcolor=blue, citecolor=blue, urlcolor=blue]{hyperref}

\newcommand{\blue}{\textcolor{blue}}

\newcommand{\hide}[1]{} 
\renewcommand{\hide}[1]{#1}

\begin{document}

\title{Momentum-Resolved Kondo Hybridization of Nonsymmorphic Dirac-Derived Bands in CeAgBi$_2$}

\author{Yang Luo}
\affiliation{Hunan Key Laboratory of Super Microstructure and Ultrafast Process, School of Physics, Central South University, Changsha 410083, Hunan, China}

\author{Hao Liu}
\affiliation{Hunan Key Laboratory of Super Microstructure and Ultrafast Process, School of Physics, Central South University, Changsha 410083, Hunan, China}

\author{Zhong-Tuo Fu}
\affiliation{Hunan Key Laboratory of Super Microstructure and Ultrafast Process, School of Physics, Central South University, Changsha 410083, Hunan, China}

\author{Qi-Yi Wu}
\affiliation{Hunan Key Laboratory of Super Microstructure and Ultrafast Process, School of Physics, Central South University, Changsha 410083, Hunan, China}

\author{Bo Chen}
\affiliation{Hunan Key Laboratory of Super Microstructure and Ultrafast Process, School of Physics, Central South University, Changsha 410083, Hunan, China}

\author{Zheng-Tai Liu}
\affiliation{Shanghai Synchrotron Radiation Facility, Shanghai Advanced Research Institute, Chinese Academy of Sciences, Shanghai 201204, China}

\author{Mao Ye}
\affiliation{Shanghai Synchrotron Radiation Facility, Shanghai Advanced Research Institute, Chinese Academy of Sciences, Shanghai 201204, China}

\author{Da-Wei Shen}
\affiliation{National Synchrotron Radiation Laboratory and School of Nuclear Science and Technology, University of Science and Technology of China, Hefei 230026, Anhui, China}

\author{Chen Zhang}
\affiliation{Hunan Key Laboratory of Super Microstructure and Ultrafast Process, School of Physics, Central South University, Changsha 410083, Hunan, China}

\author{Ying Zhou}
\affiliation{Hunan Key Laboratory of Super Microstructure and Ultrafast Process, School of Physics, Central South University, Changsha 410083, Hunan, China}

\author{Cun-Hui Lyu}
\affiliation{Hunan Key Laboratory of Super Microstructure and Ultrafast Process, School of Physics, Central South University, Changsha 410083, Hunan, China}

\author{Jin-Dong Bai}
\affiliation{Hunan Key Laboratory of Super Microstructure and Ultrafast Process, School of Physics, Central South University, Changsha 410083, Hunan, China}

\author{Zi-Jie Xu}
\affiliation{Hunan Key Laboratory of Super Microstructure and Ultrafast Process, School of Physics, Central South University, Changsha 410083, Hunan, China}

\author{Hai-Long Deng}
\affiliation{Hunan Key Laboratory of Super Microstructure and Ultrafast Process, School of Physics, Central South University, Changsha 410083, Hunan, China}

\author{Zhen-Xing Wu}
\affiliation{Hunan Key Laboratory of Super Microstructure and Ultrafast Process, School of Physics, Central South University, Changsha 410083, Hunan, China}

\author{Jun-Jie Wu}
\affiliation{Hunan Key Laboratory of Super Microstructure and Ultrafast Process, School of Physics, Central South University, Changsha 410083, Hunan, China}

\author{Sai-Chao Deng}
\affiliation{Hunan Key Laboratory of Super Microstructure and Ultrafast Process, School of Physics, Central South University, Changsha 410083, Hunan, China}

\author{Yu-Xia Duan}
\affiliation{Hunan Key Laboratory of Super Microstructure and Ultrafast Process, School of Physics, Central South University, Changsha 410083, Hunan, China}

\author{Jian-Qiao Meng}%
\email{Corresponding author: jqmeng@csu.edu.cn}
\affiliation{Hunan Key Laboratory of Super Microstructure and Ultrafast Process, School of Physics, Central South University, Changsha 410083, Hunan, China}

\date{\today}

\begin{abstract}

How symmetry-constrained conduction states acquire heavy-quasiparticle character is a central question in correlated quantum materials. Combining angle-resolved photoemission spectroscopy, Ce $4d \rightarrow 4f$ resonant photoemission, and first-principles calculations, we investigate the nonsymmorphic Kondo lattice CeAgBi$_2$. We identify quasi-two-dimensional Bi-square-net-derived bands consistent with calculated Dirac crossings and nodal lines. Resonant photoemission resolves a nearly dispersionless Ce 4$f$ band near the Fermi level, establishing Kondo heavy-quasiparticle states. Its pronounced momentum-dependent spectral weight at intersections with the identified Dirac-derived branches provides direct spectroscopic evidence for their participation in $c$–$f$ hybridization. Complementary measurements on independently cleaved samples and an effective hybridization model connect these branches to the Kondo heavy-quasiparticle spectrum. The combined results support preservation of nonsymmorphic Dirac-band connectivity under Kondo renormalization, showing how symmetry-constrained conduction states become constituents of correlated heavy quasiparticles.

\end{abstract}

\maketitle

Topological semimetals host symmetry-protected band crossings that give rise to Dirac, Weyl, and nodal-line quasiparticles, providing a platform for exploring unconventional quantum phenomena \cite{Bansil2016RMP, XWan2011PRB, SMYoung2012PRL, ZWang2013PRB, YKim2015PRL, MNeupane2014NC, ZKLiu2014S, BQLv2015NP, SYXu2015NP, SBorisenko2019NC, GBian2016NC, YZhang2025NC}. In most experimentally established systems, these quasiparticles are well described within a weakly interacting band picture \cite{MNeupane2014NC, ZKLiu2014S, BQLv2015NP, SYXu2015NP, SBorisenko2019NC, GBian2016NC, YZhang2025NC}. Kondo lattices provide a complementary setting in which hybridization between localized $f$-electron degrees of freedom and itinerant conduction states produces heavy quasiparticles and underlies unconventional superconductivity and quantum criticality \cite{Stewart1984RMP, YFYang2012PNAS, SFujimori2016JPCM, SKirchner2020RMP, MSmidman2023RMP}. Bringing these ingredients together raises a microscopic question: how do conduction states belonging to a symmetry-constrained Dirac-band manifold participate in the formation of Kondo heavy quasiparticles? Resolving this question requires identifying the parent conduction bands and tracing their coupling to the low-energy $f$-electron states, thereby connecting crystalline symmetry to the correlated quasiparticle spectrum \cite{MDzero2010PRL, PYChang2017NP, YXu2017PRX, Dzsaber2017PRL, HHLai2018PNAS, CYGuo2018NC, CCao2020PRL, LChen2022NP}.

Nonsymmorphic lattices offer a particularly useful setting for this investigation. Although spin–orbit coupling (SOC) can gap crossings protected only in the spinless limit, nonsymmorphic symmetries can enforce degeneracies in the spinful electronic structure \cite{SMYoung2015PRL}. Such crossings have been observed by angle-resolved photoemission spectroscopy (ARPES) in square-net compounds, including ZrSiS \cite{LMSchoop2016NC}, CeSbTe \cite{LMSchoop2018SA, PLi2021SCPMA}, PtPb$_4$ \cite{HWu2022NPJ}, and LaAgSb$_2$ \cite{MRosmus2022Materials}. Their coupling to correlated $f$-electron states introduces an additional level of electronic reconstruction. ARPES measurements on CeAgSb$_2$ have associated anomalies in Dirac-band dispersions with hybridization involving Ce 4$f$ states \cite{SDatta2024Nano}. A particularly revealing experimental signature is the redistribution of Kondo-resonance spectral weight at momenta where the identified Dirac-derived conduction branches intersect the near-Fermi-level $f$ band. This directly connects the parent band manifold to the heavy-quasiparticle states.

\begin{figure*}[htbp]
\includegraphics[width=\textwidth]{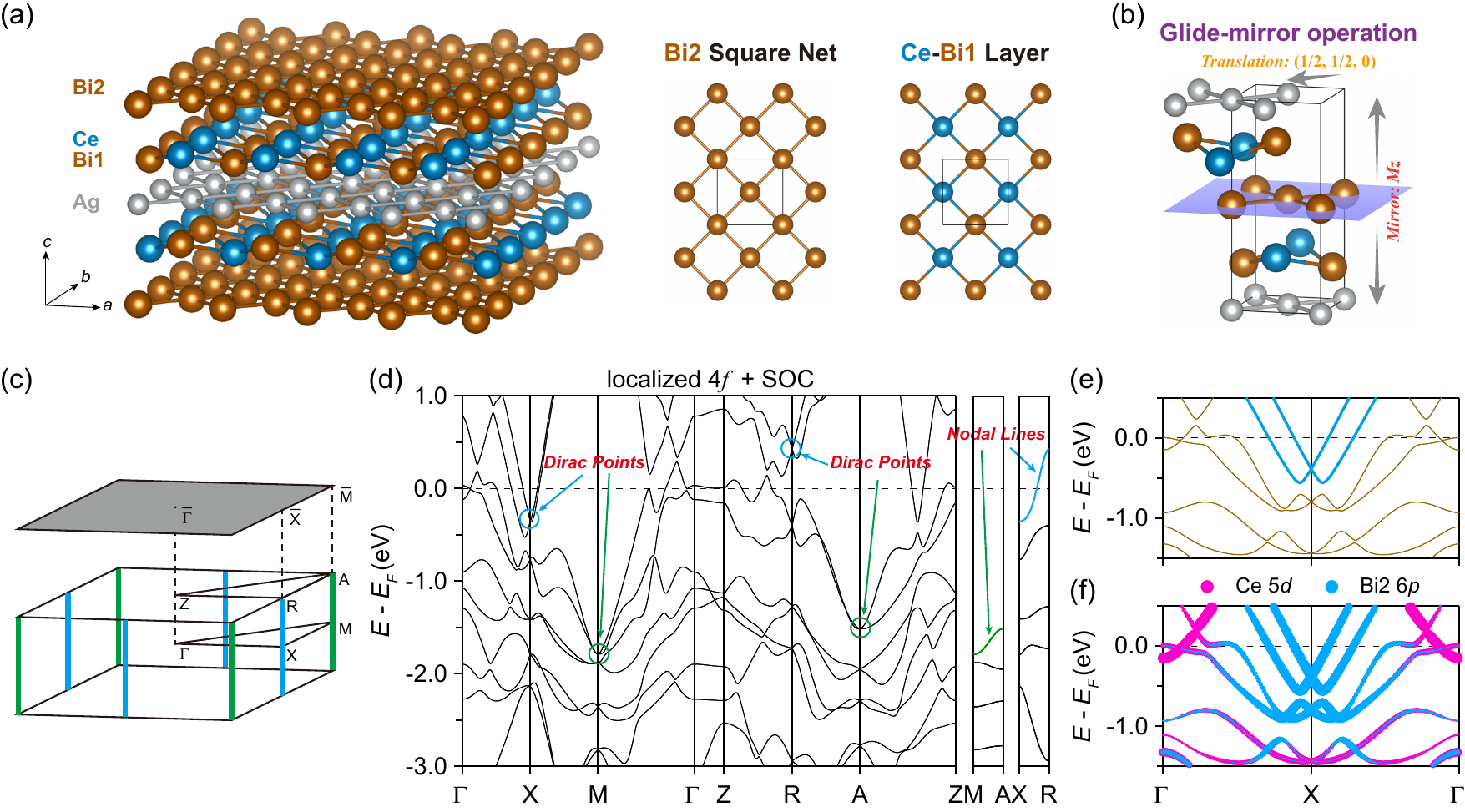}
\caption{Crystal structure and calculated electronic structure of CeAgBi$_2$. (a) Crystal structure showing the inequivalent Bi1 and Bi2 sites, which form Ce-Bi1 layers and a Bi square net, respectively. (b) Glide-mirror operation combining mirror reflection with a fractional translation. (c) Bulk and (001)-projected surface Brillouin zones. Blue and green lines indicate the calculated Dirac nodal lines. (d) Bulk band structure calculated with spin-orbit coupling and Ce $4f$ electrons treated as localized core states. Blue and green circles mark the Dirac crossings at X(R) and M(A), respectively; the corresponding bands along $X$--$R$ and $M$--$A$ are highlighted. (e) Enlarged band structure along $\Gamma$-X-$\Gamma$, with the Dirac-derived bands highlighted in blue. (f) Orbital-resolved bands along the same path, showing predominantly Bi2 $6p$ character for the Dirac-derived bands and Ce $5d$ character for the electron-like band centered at $\Gamma$.}
\label{Fig:01}
\end{figure*}

Recent theoretical studies predict Kondo-modified Dirac nodal-line states in nonsymmorphic CePt$_2$Si$_2$ \cite{HTMa2023PRB} and CeAgSb$_2$ \cite{DLGuo2025CPB}. These predictions identify Kondo hybridization as a mechanism for reshaping electronic states constrained by crystalline symmetry and motivate a band-resolved experimental examination of the participating conduction states. The layered Kondo lattice CeAgBi$_2$ combines a Bi square net (formed by Bi2 sites) with Ce 4$f$ local moments and established Kondo behavior \cite{Petrovic2003JMMM, SMThomas2016PRB, CAn2025PRB} [Figs.~\hyperref[Fig:01]{\ref{Fig:01}(a)} and \hyperref[Fig:01]{\ref{Fig:01}(b)}]. Its electronic structure therefore allows the Dirac-derived conduction bands and the near-Fermi-level Ce 4$f$ resonance to be examined within the same momentum-resolved framework.

Here, we combine ARPES, Ce $4d \rightarrow 4f$ resonant photoemission, and first-principles calculations to investigate this connection in CeAgBi$_2$. We identify quasi-two-dimensional Dirac-derived bands with predominantly Bi-square-net character. Resonant ARPES resolves a near-Fermi-level Ce 4$f$ heavy-quasiparticle band and reveals its strongly momentum-dependent spectral weight near the conduction-band intersections. The central result is the direct identification of the nonsymmorphic Dirac-derived contributions to the Kondo heavy-quasiparticle spectrum through the momentum-dependent spectral weight of the resonant Ce 4$f$ band. Measurements on independently cleaved samples and comparison with an effective periodic Anderson model (PAM) support this assignment. Combined with the experimentally observed Dirac-derived dispersions and the persistence of the calculated crossings under different treatments of the Ce 4$f$ states, these results support the coexistence of nonsymmorphic band connectivity and Kondo-induced quasiparticle renormalization.

We first establish the parent electronic structure using first-principles calculations. Figure \blue{1(c)} shows the bulk and (001)-projected Brillouin zones. With Ce 4$f$ electrons treated as localized core states and SOC included, the calculated bands exhibit Dirac crossings at X(R) and M(A) [Fig.  \blue{1(d)}], consistent with the nonsymmorphic symmetry constraints \cite{SMYoung2015PRL}. The corresponding degeneracies extend along X–R and M–A, forming the calculated Dirac nodal lines indicated in Fig. \blue{1(c)}. Their crossing energies vary along the out-of-plane direction, while their in-plane momenta remain at the Brillouin-zone boundaries. Calculations without SOC and with alternative treatments of the Ce 4$f$ states, including DFT+$U$, retain the corresponding crossings [Fig.  \blue{S2} \cite{SUPPM}]. These comparisons identify a persistent Dirac-derived conduction-band manifold that provides the parent states for examining Kondo hybridization.

\begin{figure*}[htbp]
\includegraphics[width=0.98 \textwidth]{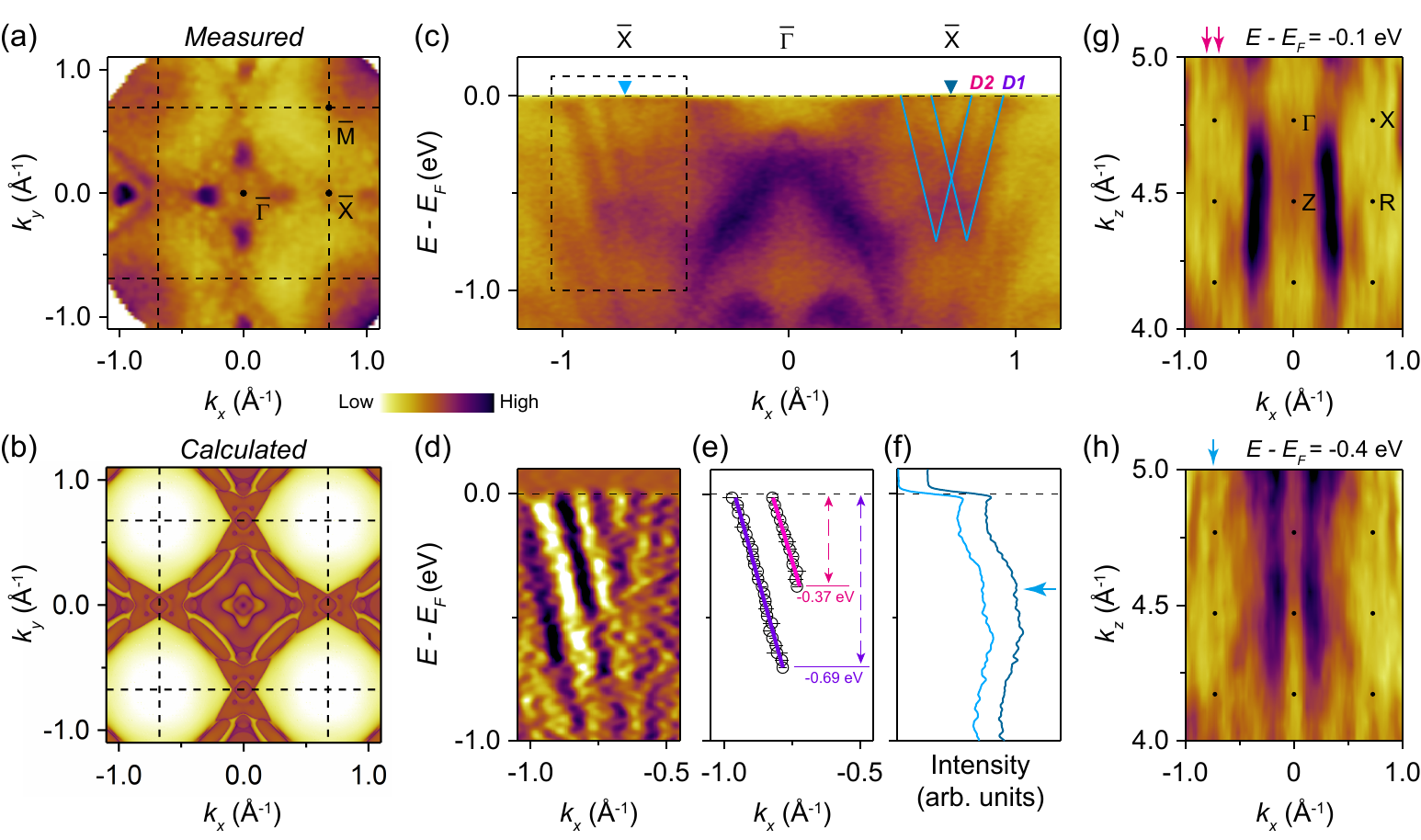}
\caption{ARPES observation of Dirac-derived bands in CeAgBi$_2$. (a) Fermi-surface map measured with 92 eV photons. (b) Surface projection of the calculated bulk Fermi surface. (c) ARPES spectrum along $\overline{\mathrm{X}}$-$\overline{\mathrm{\Gamma}}$-$\overline{\mathrm{X}}$ measured with $h\nu=92$ eV. Blue lines trace the outer (D1) and inner (D2) Dirac-derived branches. (d) Second-derivative map of the region enclosed by the dashed box in (c). (e) MDC-derived dispersions of D1 and D2 (open circles), with linear fits shown by solid lines. Error bars represent uncertainties in the extracted peak positions. (f) EDCs at the two $\overline{\mathrm{X}}$ points in (c). The blue arrow marks the crossing energy; no gap is resolved. (g),(h) Constant-energy maps in the $k_x$-$k_z$ plane at $E$-$E_F$ = -0.1 and -0.4 eV, respectively, measured over a photon-energy range of $h\nu=56$-$94$ eV in steps of $2$ eV. Arrows indicate the Dirac-derived features with weak $k_z$ dependence.}
\label{Fig:02}
\end{figure*}

The enlarged band structure along $\Gamma$-X-$\Gamma$ resolves the X-centered Dirac-derived bands [Fig. \blue{1(e)}]. Orbital projections identify predominantly Bi2 6$p$ character, associating these states with the Bi square net  [Figs. \blue{1(f)} and \blue{S3} \cite{SUPPM}]. In contrast, the electron-like band centered at $\Gamma$ has substantial Ce 5$d$ character. When the Ce 4$f$ electrons are included as itinerant valence states, Ce 4$f$ weight appears on portions of the reconstructed conduction bands [Figs. \blue{S4} and \blue{S8} \cite{SUPPM}]. The calculations thus establish the orbital basis for coupling between the Dirac-derived Bi states and the Ce 4$f$ subsystem. Because static calculations do not describe the dynamical formation of the Kondo resonance, we use resonant ARPES to determine how this coupling manifests in the low-energy quasiparticle spectrum.

The Fermi-surface map measured with 92 eV photons exhibits a diamond-shaped contour centered at $\overline{\mathrm{\Gamma}}$ [Fig. \blue{2(a)}]. Its overall geometry agrees qualitatively with the projected bulk calculation [Fig. \blue{2(b)}] and resembles that observed in related square-net compounds \cite{MRosmus2022Materials, SDatta2024Nano}. A constant-energy comparison at 0.4 eV below the Fermi level, $E_F$, provides additional support for the band assignment [Fig. \blue{S5} \cite{SUPPM}]. Along $\overline{\mathrm{X}}$-$\overline{\mathrm{\Gamma}}$-$\overline{\mathrm{X}}$, ARPES resolves steep, approximately linear branches near the projected zone boundaries [Fig. \blue{2(c)}]. We label the outer and inner Dirac-derived branches D1 and D2, respectively. Their dispersions are further resolved in the second-derivative map and extracted from momentum-distribution curves (MDCs) [Figs. \blue{2(d)} and \blue{2(e)}]. Linear fits yield band slopes $\hbar v$ of approximately 3.95 and 3.65 eV$\cdot$Å for D1 and D2, respectively, with extrapolated crossing energies near -0.69 and -0.37 eV. The energy-distribution curves (EDCs) at the two projected $\overline{\mathrm{X}}$ points resolve the crossing feature indicated in Fig. \blue{2(f)}. The measured band geometry, linear dispersions, and calculated orbital character together identify the Dirac-derived conduction states whose low-energy hybridization is examined below.

\begin{figure*}[htbp]
\includegraphics[width=0.82\textwidth]{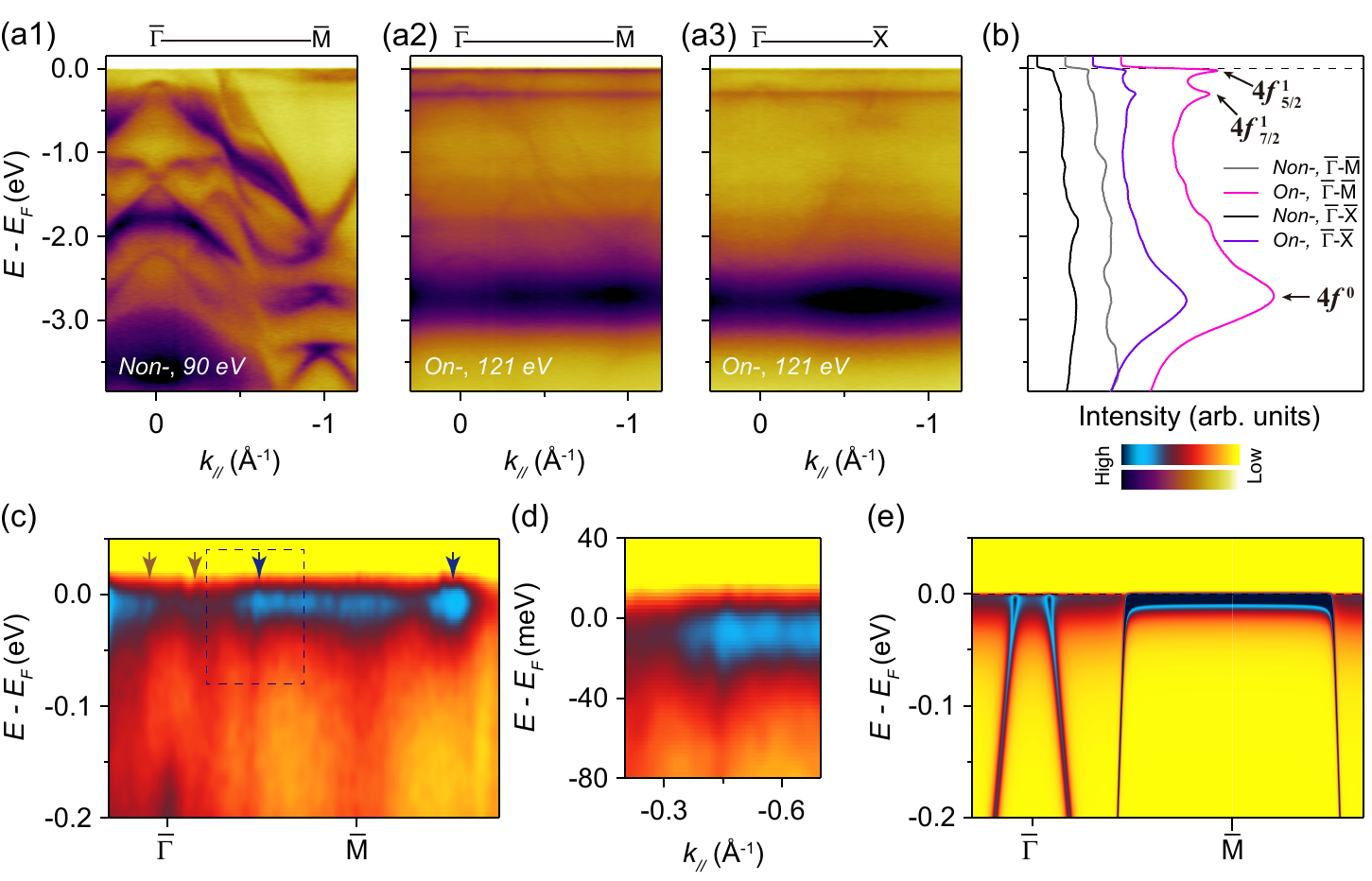}
\caption{Resonant ARPES signatures of Kondo hybridization in CeAgBi$_2$. (a1) Nonresonant spectrum along $\overline{\mathrm{\Gamma}}$-$\overline{\mathrm{M}}$ measured with $h\nu$ = 90 eV. (a2),(a3) On-resonance spectra along $\overline{\mathrm{\Gamma}}$-$\overline{\mathrm{M}}$ and $\overline{\mathrm{\Gamma}}$-$\overline{\mathrm{X}}$, respectively, measured with $h\nu$=121 eV at the Ce $4d\rightarrow4f$ resonance. (b) Angle-integrated EDCs for the two momentum cuts measured with nonresonant and on-resonance photons, showing resonant enhancement of the $4f^{1}_{5/2}$ Kondo resonance, its $4f^{1}_{7/2}$ spin-orbit satellite, and the $4f^0$ final-state feature. The momentum integration windows span distances equivalent to the $\overline{\mathrm{\Gamma}}$-$\overline{\mathrm{M}}$ and $\overline{\mathrm{\Gamma}}$-$\overline{\mathrm{X}}$ distances, respectively. (c) On-resonance spectrum near the Fermi level along $\overline{\mathrm{\Gamma}}$-$\overline{\mathrm{M}}$, showing conduction-band bending and momentum-dependent redistribution of the $4f^{1}_{5/2}$ spectral weight. Arrows mark the hybridization regions. (d) An enlarged view of the hybridization region indicated by the dashed box in (c). (e) Hybridized spectral-intensity map calculated using the effective hybridization model [Eq.~(\blue{1})], with $V_1$=15 meV for the electron-like band near $\overline{\mathrm{\Gamma}}$ and $V_2$=30 meV for the hole-like band approaching $\overline{\mathrm{M}}$. }
\label{Fig:03}
\end{figure*}

Photon-energy-dependent measurements probe the out-of-plane evolution of the Dirac-derived bands. Constant-energy maps at 0.1 and 0.4 eV below $E_F$, acquired over a photon-energy range of 56–94 eV, show weak photon-energy dependence of the corresponding features [Figs. \blue{2(g)} and \blue{2(h)}]. Additional spectra measured at 81, 90, and 96 eV retain similar dispersions and crossing features [Fig. \blue{S6} \cite{SUPPM}]. Together with the measured band geometry and its correspondence with the projected bulk calculations, these observations support the assignment of the branches to the quasi-two-dimensional Bi-square-net-derived Dirac-band manifold and are consistent with the calculated nodal-line structure. Having identified the parent Dirac-derived bands, we next examine their connection to the Ce 4$f$ heavy-quasiparticle states near $E_F$.

\begin{figure*}[htbp]
\includegraphics[width=\textwidth]{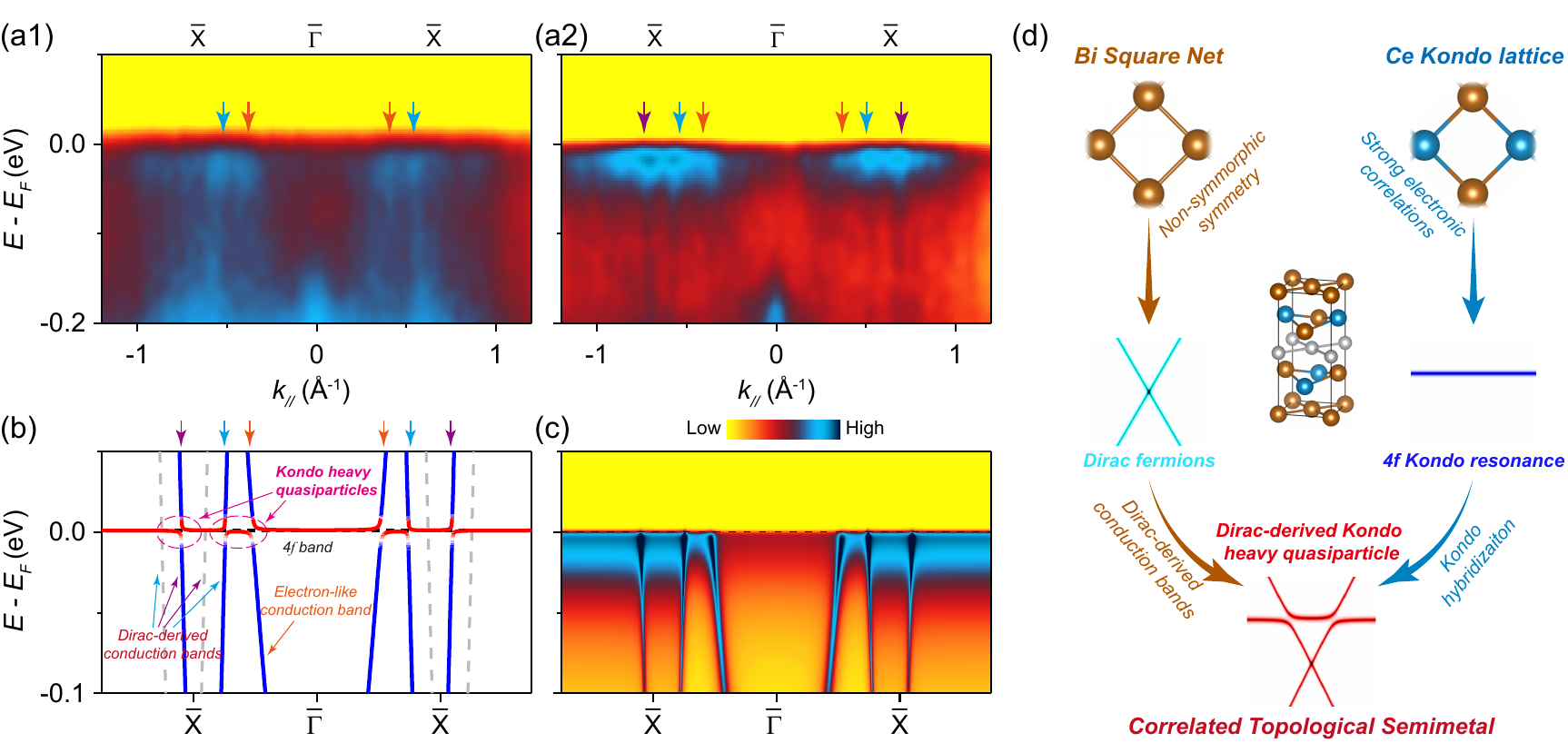}
\caption{Kondo hybridization of Dirac-derived bands in CeAgBi$_2$. (a1),(a2) On-resonance ARPES spectra along $\overline{\mathrm{X}}$-$\overline{\mathrm{\Gamma}}$-$\overline{\mathrm{X}}$ measured with $h\nu$=121 eV on two independently cleaved samples, S1 and S2. Arrows indicate the D1 and D2 Dirac-derived branches and the electron-like band centered at $\overline{\mathrm{\Gamma}}$. Significant momentum-dependent redistribution of the Ce $4f^{1}_{5/2}$ spectral weight is observed near the Dirac-derived branches. (b),(c) Hybridized dispersions and corresponding spectral-intensity map calculated using the effective hybridization model, including D1, D2, the $\Gamma$-centered electron-like band, and the Ce $4f$ resonance. Dashed lines in (b) indicate the uncoupled bands; dashed circles mark the hybridization regions. (d) Schematic illustration of $c$-$f$ hybridization between the Bi-square-net-derived Dirac bands and the Ce $4f$ states, showing the coexistence of Dirac-band connectivity and Kondo-induced quasiparticle reconstruction. }
\label{Fig:04}
\end{figure*}

Ce $4d \rightarrow 4f$ resonant photoemission identifies the Ce-derived contributions to the electronic spectrum. The nonresonant spectrum measured with 90 eV photons is dominated by dispersive conduction bands [Fig. \blue{3(a1)}]. At 121 eV, the spectra along $\overline{\mathrm{\Gamma}}$-$\overline{\mathrm{M}}$ and $\overline{\mathrm{\Gamma}}$-$\overline{\mathrm{X}}$ exhibit enhanced Ce 4$f$ emission [Figs. \blue{3(a2)} and \blue{3(a3)}]. The angle-integrated EDCs resolve the near-$E_F$ $4f^1_{5/2}$ resonance, its $4f^1_{7/2}$ spin–orbit satellite near -0.3 eV, and the $4f^0$ final-state feature near -2.75 eV [Fig. \blue{3(b)}], characteristic of Ce-based Kondo systems \cite{SPatil2016NC, QYChen2017PRB, YLuo2020PRB, ZWu2021PRL, Chen2024PRB, Chen2024PRX, Song2021CPL}. The nearly dispersionless $4f^1_{5/2}$ band near $E_F$ identifies the Kondo heavy-quasiparticle states in CeAgBi$_2$. Its momentum-resolved spectral weight provides a direct means of determining which conduction bands participate in their formation.

The enlarged resonant spectrum along $\overline{\mathrm{\Gamma}}$-$\overline{\mathrm{M}}$ reveals conduction-band bending near the heavy-quasiparticle band [Figs. \blue{3(c)} and \blue{3(d)}]. The Ce 4$f$ spectral weight is strongly momentum dependent, with pronounced redistribution near the conduction-band intersections. This combination of a near-$E_F$ heavy-quasiparticle band, conduction-band reconstruction, and intersection-associated 4$f$ spectral weight constitutes the spectroscopic signature of $c$–$f$ hybridization \cite{Song2021CPL, YZZhao2024, JJSong2024}. It establishes the low-energy coupling between the Ce 4$f$ states and the dispersive conduction bands and motivates a corresponding examination of the Dirac-derived branches.

To describe this reconstruction, we use an effective low-energy hybridization model motivated by the PAM \cite{QYChen2017PRB, ACHewson1993}. Each conduction-band–$f$-level intersection is described locally by coupling a bare dispersion $\varepsilon_i(k)$ to a nearly dispersionless effective f level $\varepsilon_f$. The hybridized branches are
\begin{equation}
E_{i,\pm}(k)=
\frac{\varepsilon_f+\varepsilon_i(k)\pm\sqrt{[\varepsilon_f-\varepsilon_i(k)]^2+4|V_i|^2}}{2},
\label{eq:PAM}
\end{equation}
where $V_i$ is the effective hybridization amplitude for conduction band $i$. The model uses $V_1$=15 meV for the electron-like band near $\overline{\mathrm{\Gamma}}$ and $V_2$=30 meV for the hole-like band approaching $\overline{\mathrm{M}}$. The resulting dispersions and intensity map capture the principal reconstruction observed along $\overline{\mathrm{\Gamma}}$–$\overline{\mathrm{M}}$ [Figs. \blue{3(e)} and \blue{S7} \cite{SUPPM}]. As a conduction band approaches the effective 4$f$ level, hybridization bends its dispersion and transfers $f$-electron character to the resulting quasiparticle branches. The momentum dependence of this mixing connects the observed band reconstruction with the redistribution of resonant spectral weight. The parameters serve as a phenomenological description of the local hybridization regions.

We next directly examine whether the Dirac-derived conduction states participate in Kondo hybridization. Ce $4d \rightarrow 4f$ resonant ARPES measurements were performed along $\overline{\mathrm{X}}$-$\overline{\mathrm{\Gamma}}$-$\overline{\mathrm{X}}$, which intersects the upper branches of the $\overline{\mathrm{X}}$-centered Dirac bands as they approach $E_F$. Figures \blue{4(a1)} and \blue{4(a2)} show spectra acquired on two independently cleaved samples, S1 and S2. Both measurements resolve the nearly dispersionless Ce 4$f$ Kondo heavy-quasiparticle band near $E_F$, although the relative visibility of the conduction branches differs. In S1, the outer Dirac-derived branch D1 exhibits pronounced enhancement of the near-$E_F$ Ce 4$f$ spectral weight, whereas in S2, a corresponding enhancement occurs near the inner branch D2. The strong momentum dependence of the heavy-quasiparticle spectral weight at the intersections with the identified Dirac-derived branches provides direct spectroscopic evidence for their participation in $c$–$f$ hybridization. The complementary observations on the two cleaves establish that the Dirac-derived conduction states contribute to the Kondo-hybridized low-energy spectrum.

To describe the resulting electronic reconstruction, we apply the same effective hybridization framework to the $\overline{\mathrm{X}}$-$\overline{\mathrm{\Gamma}}$-$\overline{\mathrm{X}}$ spectra. The calculated dispersions in Fig. \blue{4(b)} include the Dirac-derived branches, the $\overline{\mathrm{\Gamma}}$-centered electron-like band, and the nearly dispersionless Ce 4$f$ states. The corresponding intensity map in Fig. \blue{4(c)} captures the principal momentum-dependent features of the experimental spectra. The agreement associates the enhanced Ce 4$f$ spectral weight with hybridized quasiparticles carrying both Dirac-derived conduction-band and $f$-electron character. Together with the resonant measurements, these calculations establish a microscopic picture in which the nonsymmorphic Dirac-derived states actively participate in the formation of Kondo heavy quasiparticles.

An important aspect of CeAgBi$_2$ is the coexistence of nonsymmorphic symmetry protection and Kondo-induced quasiparticle renormalization. The corresponding Dirac crossings persist in calculations including SOC and different treatments of the Ce 4$f$ electrons [Figs. \blue{1(d)} and \blue{S2} \cite{SUPPM}]. ARPES resolves the associated Dirac-derived dispersions without a resolved gap at the observed crossing feature, while their weak photon-energy dependence supports the quasi-two-dimensional electronic structure associated with the Bi square net. Taken together with the resonant hybridization signatures, this correspondence between experiment and calculation supports the preservation of the underlying Dirac-band connectivity as the low-energy states acquire Kondo heavy-quasiparticle character. Crystalline symmetry constrains the band degeneracies and connectivity, whereas $c$–$f$ hybridization reshapes the quasiparticle dispersions and orbital composition.

This coexistence has a natural microscopic basis in the orbital structure of CeAgBi$_2$. The Bi square net supplies the Dirac-derived conduction bands, while the Ce 4$f$ degrees of freedom generate the near-$E_F$ Kondo resonance through hybridization with itinerant states. The resulting heavy quasiparticles inherit the conduction-band character of the symmetry-constrained Dirac manifold, linking the spatially distinct Bi and Ce electronic subsystems through $c$–$f$ mixing [Fig. \blue{4(d)}]. Symmetry protection constrains the allowed band crossings without fixing the quasiparticle velocity or orbital composition. It can therefore accommodate substantial many-body renormalization. CeAgBi$_2$ illustrates how a nonsymmorphic electronic structure can retain its characteristic band connectivity while developing strongly correlated low-energy states.

In EuRh$_2$Si$_2$, linear conduction bands interact with localized Eu 4$f$ excitations \cite{Hoppner20132013NC}, while Dirac-band anomalies associated with Ce 4$f$ hybridization have been reported in CeAgSb$_2$ \cite{SDatta2024Nano}. Here, resonant ARPES connects identified nonsymmorphic Dirac-derived conduction branches directly to a near-$E_F$ Ce Kondo heavy-quasiparticle band through its strongly momentum-dependent spectral weight. This connection provides an experimental counterpart to theoretical proposals of Kondo-modified Dirac nodal-line states in nonsymmorphic systems \cite{HTMa2023PRB, DLGuo2025CPB}. More broadly, our results show how symmetry-constrained Dirac-derived states can become constituents of a correlated heavy-quasiparticle spectrum, providing a platform for exploring the interplay of crystalline symmetry, Kondo hybridization, and nodal electronic structure.

In summary, we have provided direct momentum-resolved spectroscopic evidence that nonsymmorphic Dirac-derived conduction states participate in Kondo hybridization, forming Dirac-derived heavy quasiparticles in CeAgBi$_2$. First-principles calculations identify nonsymmorphic symmetry-protected Dirac crossings and nodal lines, while ARPES resolves the corresponding quasi-two-dimensional Bi-square-net-derived bands. Resonant photoemission reveals the near-$E_F$ Ce 4$f$ heavy-quasiparticle band, whose pronounced spectral-weight enhancement at intersections with the identified Dirac-derived branches provides direct evidence for their hybridization. The combined experimental and theoretical results support the preservation of the underlying Dirac-band connectivity while the low-energy states acquire Kondo heavy-quasiparticle character. These findings show how crystalline-symmetry constraints and many-body hybridization jointly shape the electronic structure, providing an experimental basis for exploring correlated Dirac-derived quasiparticles in $f$-electron materials.

This work was supported by the National Natural Science Foundation of China (Grant Nos. 12604111, 12574168), the National Key Research and Development Program of China (Grant No. 2022YFA1604204). Part of this work was carried out using the computing resources at the High Performance Computing Center of Central South University.

$Data$ $availability$—The data that support the findings of this article are not publicly available. The data are available from the authors upon reasonable request.

\end{document}